\documentclass[a4paper,11pt]{article}
\usepackage{pos}
\usepackage[font=footnotesize]{caption}
\title{Analysis of commissioning data from SST-1M : A Prototype of Single-Mirror Small Size Telescope}
 \ShortTitle{SST-1M commissioning data analysis}

\author*[a]{T.~Tavernier}
\author[a]{J.~Jury\v{s}ek}
\author[a, b]{V.~Novotn\'{y}}
\author[c]{M.~Heller}
\author[a]{D.~Mandat}
\author[a]{M.~Pech}
\author[c]{C.~Alispach}
\author[d,e]{A.~Araudo}
\author[f]{V.~Beshley}
\author[a]{J.~Blazek}
\author[g]{J.~Borkowski}
\author[h]{S.~Boula}
\author[i]{T.~Bulik}
\author[c]{F.~Cadoux}
\author[h]{S.~Casanova}
\author[a]{A.~Christov}
\author[j]{L.~Chytka}
\author[c]{D.~della Volpe}
\author[c]{Y.~Favre}
\author[k]{L.~Gibaud}
\author[h]{T.~Gieras}
\author[j]{P.~Hamal}
\author[j]{M.~Hrabovsky}
\author[l]{M.~Jel\'inek}
\author[d]{V.~Karas}
\author[m]{E.~Lyard}
\author[h]{E.~Mach}
\author[h]{W.~Marek}
\author[j]{S.~Michal}
\author[h]{J.~Micha{\l}owski}
\author[g]{R.~Moderski}
\author[c]{T.~Montaruli}
\author[g]{A.~Muraczewski}
\author[a]{S.~R.~Muthyala}
\author[c]{A.~Nagai}
\author[h]{K.~Nalewajski}
\author[n]{D.~Neise}
\author[h]{J.~Niemiec}
\author[k]{M.~Niko{\l}ajuk}
\author[o]{M.~Ostrowski}
\author[a]{M.~Palatka}
\author[a]{M.~Prouza}
\author[p]{P.~Rajda}
\author[a]{P.~Schovanek}
\author[q]{K.~Seweryn}
\author[m]{V.~Sliusar}
\author[o]{{\L}.~Stawarz}
\author[h]{J.~\'{S}wierblewski}
\author[h]{P.~\'{S}wierk}
\author[l]{J.~\v{S}trobl}
\author[a]{J.~V\'icha}
\author[m]{R.~Walter}
\author[o]{A.~Zagda\'{n}ski}
\author[o]{K.~Zi{\c e}tara}

\scriptsize
\affiliation[]{
$^a$FZU - Institute of Physics of the Czech Academy of Sciences, Na Slovance 1999/2, Prague 8, Czech Republic
$^{b}$Charles University, Faculty of Mathematics and Physics, Institute of Particle and Nuclear Physics, Prague, Czech Republic
$^c$DPNC - Universit\'e de Gen\`eve, 24 Quai Ernest Ansermet, CH-1211 Gen\`eve,  Switzerland
$^d$Astronomical Institute of the Czech Academy of Sciences, Bo\v{c}n\'i II 1401, CZ-14100 Prague, Czech Republic
$^e$ELI Beamlines, Institute of Physics, Czech Academy of Sciences, CZ-25241 Doln\'i B\v{r}e\v{z}any, Czech Republic
$^f$Pidstryhach Institute for Applied Problems of Mechanics and Mathematics, National Academy of Sciences of Ukraine, 3-b Naukova St., 79060, Lviv, Ukraine
$^g$Nicolaus Copernicus Astronomical Center, Polish Academy of Sciences,  ul. Bartycka 18, 00-716 Warsaw, Poland
$^h$Institute of Nuclear Physics Polish Academy of Sciences, PL-31342 Krakow, Poland
$^i$Astronomical Observatory, University of Warsaw, Al. Ujazdowskie 4, 00-478 Warsaw, Poland
$^j$Palacky University Olomouc, Faculty of Science, 17. listopadu 50, Olomouc, Czech Republic
$^{k}$Faculty of Physics, University of Bia{\l}ystok, ul. K. Cio{\l}kowskiego 1L, 15-245 Bia{\l}ystok, Poland
$^{l}$Astronomical Institute of the Czech Academy of Sciences, Fri\v{c}ova 298, CZ-25165 Ond\v{r}ejov, Czech Republic
$^{m}$D\'epartement d'Astronomie, Universit\'e de Gen\`eve, Chemin d'Ecogia 16, CH-1290 Versoix, Switzerland
$^{n}$ETH Zurich, Institute for Particle Physics and Astrophysics, Otto-Stern-Weg 5, 8093 Zurich, Switzerland
$^{o}$Astronomical Observatory, Jagiellonian University, ul. Orla 171, 30-244 Krakow, Poland
$^{p}$AGH University of Science and Technology, al.Mickiewicza 30,  30-059 Krakow, Poland
$^{q}$Centrum Bada{\'n} Kosmicznych Polskiej Akademii Nauk,  18a Bartycka str., 00-716 Warsaw, Poland}

\emailAdd{tavernier@fzu.cz}

\abstract{SST-1M is a prototype of a single-mirror Small Size Telescope developed by a consortium of institutes from Poland, Switzerland and the Czech Republic. With a wide field of view of 9 degrees, SST-1Ms are designed to detect gamma-rays in the energy range between 1 and 300 TeV. The design of the SST-1M follows the Davies-Cotton concept, with a 9.42m2 multi-segment mirror. 
SST-1M is equipped with DigiCam camera, which features a fully digital readout and trigger system using 250 MHz ADC, and a compact Photo-Detector Plane (PDP) composed of 1296 pixels, each made of a hexagonal light guide coupled to silicone photomultipliers (SiPM).

Two SST-1M telescopes are currently being commissioned at the Ondrejov Observatory in the Czech Republic, where they are successfully observing Cerenkov events in stereo.
This contribution will present an overview of calibration strategies and performance evaluation based on data collected at the observatory.}

\ConferenceLogo{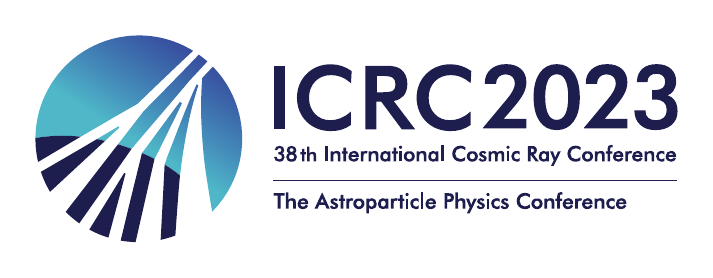}

\FullConference{%
The 38th International Cosmic Ray Conference (ICRC2023)\\
  26 July -- 3 August, 2023\\
  Nagoya, Japan}

\begin{document}
\maketitle

\section{Introduction}

The SST-1M telescope was initially developed as a prototype for a Small-Sized Telescope intended for the Cherenkov Telescope Array (CTA). Its primary purpose, when deployed in an array of telescopes, is to observe gamma-ray induced atmospheric showers with energies between 1 and 300 TeV. The design of the SST-1M follows the Davies-Cotton concept, with a focal length of 5.6 m. The mirror of the telescope comprises 18 hexagonal facets, each measuring 78 cm (flat-to-flat), with a spherical shape and a radius of curvature of 11.2 m. This configuration maximizes the mirror area while maintaining a required point spread function (psf) below 0.25 degrees. The corrected mirror area is 9.42 m2, with a psf of 0.09 degree on-axis and 0.21 degrees at a 4-degree off-axis angle. The mirror's collecting area, accounting for shadowing, is 7.6~m$^2$. The optical time spread is less than 0.84 ns (rms) \cite{cta_sst1m, moderski_4m_2013}.

This telescope incorporates an innovative camera known as the DigiCam \cite{cta_sst1m} which include a fully digital readout and trigger system. It utilizes a 250 MHz analog-to-digital converter (ADC) for high-speed signal processing. The Photo-Detector Plane (PDP) of the camera consists of 1296 pixels, each comprising a hexagonal light guide coupled to silicone photomultipliers (SiPMs). The SiPM technology enables the camera to operate effectively even under high Night Sky Background (NSB) conditions, significantly increasing the telescope's duty cycle. Furthermore, the fully digital trigger running in real-time in FPGAs enables a flexible and adaptive system capable of implementing various triggering schemes running in parallel and dynamic per-pixel thresholds.

In 2022, a couple of SST-1M telescopes was installed (with a relative distance of 155 meters) at the Ond\v{r}ejov Observatory in the Czech Republic. At this location, the prototypes are being commissioned and tested under astronomical conditions to evaluate their capabilities for both mono and stereo observations. % The final decision regarding their ultimate deployment site is yet to be determined.

In the following sections, we describe the calibration procedures used on site to determine the instrument's response. Two different approaches are described : one involves the analysis of the single photo-electron spectrum observed during dark runs, the other one takes advantage of the signal emitted by cosmic muons. Additionally, we present first preliminary results from the Crab nebula stereo observations of SST-1M telescopes. 

%which yield significant observations regarding the instrument's performance and its inherent capabilities.

\section{Calibration}
\subsection{Single photo-electron spectrum}
\label{calib}
The gain of a SiPM coupled to the readout chain, down to the FADC, is quantified as the number of ADC counts recorded per photo-electron (p.e.).
Optical cross-talk occurs when a secondary avalanche is initiated within a micro-cell of the SiPM due to the emission of photons during the discharge process of a primary cell \cite{nagai_characterization_2019}. This phenomenon only occurs between micro-cell in the same pixels and leads to an overestimation of the detected photon count. To correct this effect and accurately measure the number of detected photons, it is essential to determine the crosstalk probability. 

On-site determination of both the gain and crosstalk probability can be achieved by analyzing multiple photo-electron spectra obtained during dedicated dark runs. During these dark runs, the SiPM is not exposed to any light source, but thermal photons might induce avalanches in the SiPM. The multiple photo-electron spectrum refers to the distribution of integrated ADC counts from  15 successive samples randomly selected within the readout window. The probability of measuring $n$ photo-electrons in a random window is given by the Generalized Poisson distribution \cite{alispach_large_2020} : 
\begin{equation}
P_{\theta,\mu}(n) = e^{-\theta - n\mu}\times\frac{\theta(\theta + n\mu)^{n-1}}{n!}
\end{equation}

which takes into account both the Poisson probability from black body radiation with $E = \theta$ and the crosstalk probability $\mu$.

Assuming both the electronic noise  and the single photo-electron response follow Gaussian fluctuation with standard deviation respectively $\sigma_\mathrm{el}$ and $\sigma_\mathrm{pe}$, the multiple photo-electron spectrum can be described by : 
\begin{equation}
\mathrm{mes}(n_\mathrm{ADC}) = \sum\limits_{n=0}^{\infty} P_{\theta,\mu}(n) \times \mathrm{Gauss}\left( n \times g,\sqrt{\sigma_\mathrm{el}^2 + g \times n \times \sigma_\mathrm{pe}^2 }\right) (n_\mathrm{ADC}) 
\label{eq:mes}
\end{equation}
where $g$ is the SiPM gain. The factor to convert integrated ADC count to p.e. is then given by $g^* = \frac{1-\mu}{g}$.

\begin{figure}[!h]
\centering
\begin{tabular}{c}
\includegraphics[width=.8\textwidth]{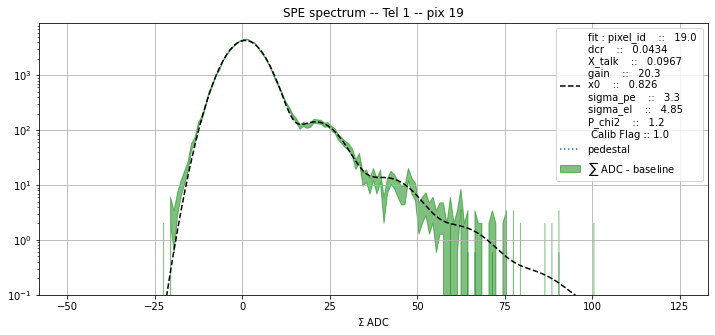}
\end{tabular}
\caption{Multiple photo-electron spectra derived from dark runs. Green band shows 1$\sigma$ confidence level of integrated ADC counts distribution. Dotted line represents the best fit obtained.}
\label{fig:mes}
\end{figure}

\begin{figure}[!htb]
\minipage{0.32\textwidth}
  \includegraphics[width=\linewidth]{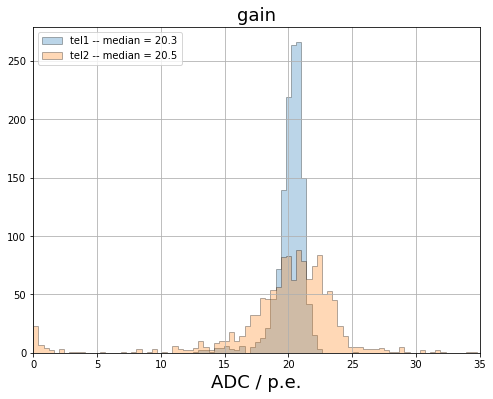}

\endminipage\hfill
\minipage{0.32\textwidth}
  \includegraphics[width=\linewidth]{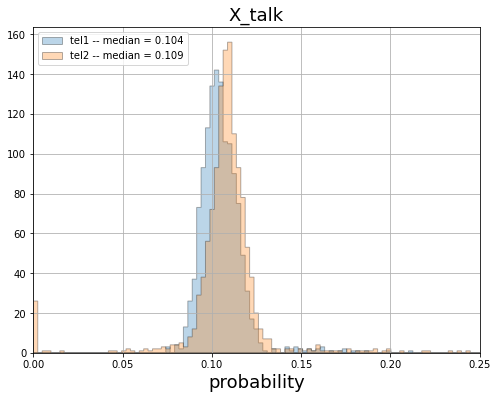}

\endminipage\hfill
\minipage{0.32\textwidth}%
  \includegraphics[width=\linewidth]{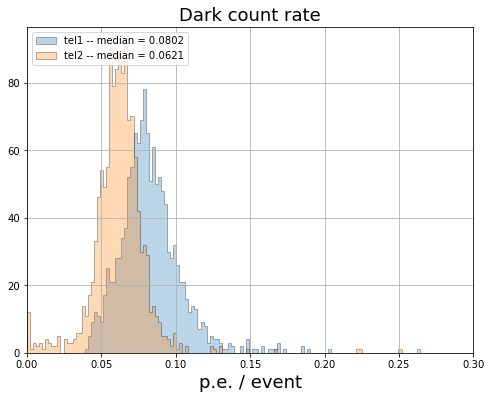}

\endminipage
\caption{Distribution of SiPM's gain (\textit{left}), X-talk (\textit{center}) and dark count rate (\textit{right}) for all pixels in both SST-1M-1 (blue)  and SST-1M-2 (orange).}
\label{fig:param_dist}
\end{figure}

In practice, the multiple photo-electron distribution is fitted with function (\ref{eq:mes}) to extract the gain, the crosstalk probability and other relevant parameters of the SiPM response. An example of such distribution for a single pixel in the camera of SST-1M-1 and the corresponding fit is shown in Figure \ref{fig:mes}. The values of the parameters obtained with the fit are used to calibrate camera images as described in section \ref{ana}.
During the commissioning observation period, dark runs are performed at the beginning and end of each observation night to monitor the stability of the SiPM's response over both short-term and long-term timescales.
The distributions of some relevant fitted parameters for both telescopes are shown in Figure \ref{fig:param_dist}.
In the camera of SST-1M-2, there are a few pixels (approximately 20-30) that exhibit a lack of gain, indicating that they are not functioning properly. The camera is undergoing maintenance to address this issue.

\subsection{Muon analysis}
\label{muon}
Cosmic muons are known to produce typical ring images at the focal plane of IACT and the number of photons hitting the detector can be estimated analytically. The nice properties of these images make muons an excellent test beam for instrument calibration and they have been used for this purpose since the early 90’s by the Whipple collaboration \cite{fleury_cerenkov_1991}. The studies presented here mostly follow procedures and recommendations given in \cite{gaug_using_2019-2}.

The selection of muon events for analysis is performed using the data accumulated during normal observation. Initially, the image is subjected to a cleaning process using the tailcuts method, which involves applying two thresholds ( 5 p.e. and 4 p.e. for the image threshold and the boundary threshold, respectively) and specifying a minimum of two neighboring pixels. Subsequently, a circle is fitted to the remaining cleaned pixels, and any pixels with their centers located within a distance of 0.15 degrees from the fitted circle are retained for further analysis.

To evaluate the completeness of the muon ring, the signal within the ring region is integrated across 12 bins of 30° angular section. In order to ensure the selection of well-defined ring structures, only images that exhibit a signal above 7 p.e. in at least 8 out of the 12 bins are considered for further analysis. 
Additionally, the signal outside the selected pixels is also taken into account. Images with an integrated charge exceeding 20 p.e. outside the selected pixels are excluded from the analysis. Both of these criteria serve to eliminate shower images or noise fluctuations, ensuring that only clean muon images with a sufficiently clear ring structure are included in the calibration studies.
The rate of muon images surviving selection cuts is about 0.1 Hz for a typical observation.

Following results are derived from data obtained during the Crab observation campaign conducted in March 2023, slightly exceeding 10 hours of data taking with zenithal angle from 35° to 65°. The total number of muon images used in this study is respectively 4311 and 5217 for SST-1M-1 and SST-1M-2. The following analysis is also run through Monte-Carlo muon images produced in \texttt{CORSIKA} \cite{1998cmcc.book.....H} and \texttt{sim\_telarray} \cite{BERNLOHR2008149} framework.

\begin{figure}[!htb]
\minipage{0.45\textwidth}
  \includegraphics[width=\linewidth]{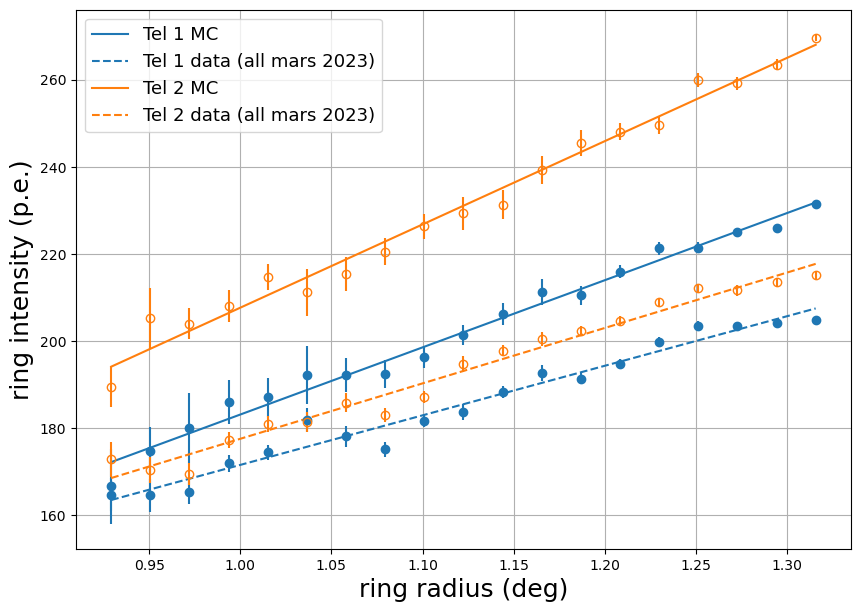}

\endminipage\hfill
\minipage{0.45\textwidth}
  \includegraphics[width=\linewidth]{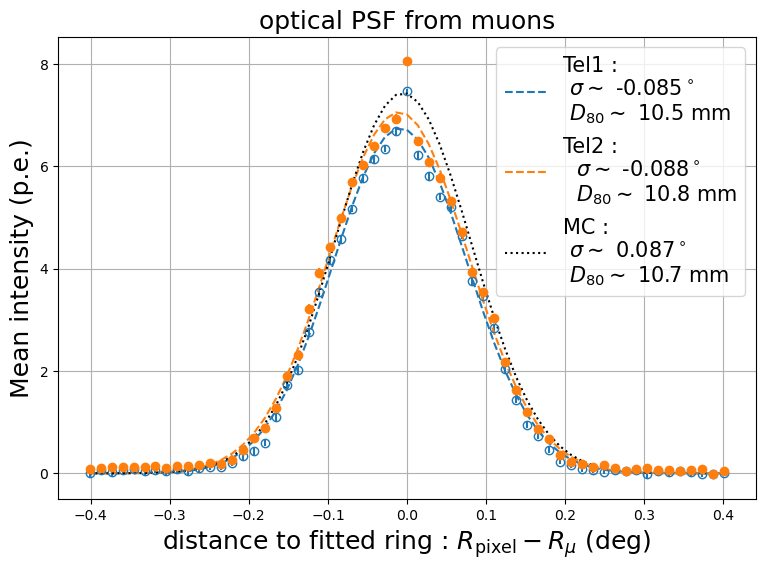}

\endminipage
\caption{\textit{Left} : Integrated muon charge as a function of muon ring radius. 
The figure shows the linear fits of the observed data (dashed line) and Monte Carlo simulation (solid line) for both SST-1M-1 (blue) and SST-1M-2 (orange). \\
\textit{Right} : Average pixel intensity plotted against the radial distance from the center of the pixel to the fitted circle. The distributions are fitted with Gaussian functions (dashed lines). SST-1M-1 is represented by the blue curve, SST-1M-2 by the orange curve, and the Monte Carlo simulation is shown in black.}
\label{fig.muon_lin}
\label{fig.mu_psf}
\end{figure}

%We present the results obtained from the analysis of muon images accumulated during the observation campaign in March 2023. These images were used to evaluate the optical throughput of the telescopes. 
The intensity of the muon rings, obtained by integrating the signal from all pixels within the muon ring region, was plotted against their respective radii. The results, along with the corresponding linear fits, are shown in Figure \ref{fig.muon_lin} for both telescopes, including both real and simulated data.

A notable discrepancy is visible between telescopes for both the real instrument data and the Monte Carlo simulations. This discrepancy can be attributed to the different transmission properties of the transmission window mounted on each telescope.

The optical efficiency was found to be 10\% (SST1-M-1) to 20\% (SST1-M-2) higher in the Monte Carlo simulations compared to the real instrument data. Further investigation revealed that the discrepancy in optical efficiency is likely induced by inaccurate consideration of mirror reflectivity in the Monte Carlo simulations. This indicates the need for improvements in the modeling of mirror reflectivity to achieve better agreement between the Monte Carlo simulations and the actual instrument's performance. Furthermore, this observed difference is expected to introduce significant bias in the energy reconstruction within the current stage of the analysis pipeline. This bias may also result in discrepancies in energy thresholds between the real instrument and its Monte-Carlo model.

%The higher optical efficiency in the Monte Carlo simulations compared to the real instrument is attributed to the inaccurate simulation of mirror reflectivity. This discrepancy will be addressed and corrected in the upcoming Monte Carlo production.

The second study using muon images is the estimation of the optical psf. This is done looking at the intensity of the signal in each pixel against its radial distance from the center of the pixel to the fitted circle. This distribution is well described by a Gaussian function which is used to determine the optical psf. Figure \ref{fig.mu_psf} illustrates this distribution for both telescopes, along with the corresponding Gaussian fits for both the real data and the Monte Carlo simulations.

D$_{80}$ is defined as the 80\,\% containment radius in the camera frame of all signal comprised in a disk of 16.2 mm radius and correspond to the definition used from estimation based on star observations. From the analysis of muon data, the estimated D$_{80}$ values for SST-1M-1 and SST-1M-2 are 10.5 $\pm$ 0.1 mm (stat) and 10.8 $\pm$ 0.1 mm (stat), respectively. These results show a very good agreement with the values obtained from psf estimation based on star observations, which are 9.6 mm and 10.8 mm for SST-1M-1 and SST-1M-2, respectively.

%\subsubsection{Point spread function}
\section{Crab analysis}
\subsection{Data acquisition}
The SST-1M Crab observation campaign in 2023 marked the first deployment of both telescopes for simultaneous data acquisition. It is important to note that at the time of this campaign, the stereo trigger implementation was still pending. Consequently, stereo events were reconstructed based on their respective time-stamps rather than relying on a dedicated stereo trigger.

\begin{figure}[h!]
\centering
\begin{tabular}{c}
\includegraphics[width=.8\textwidth]{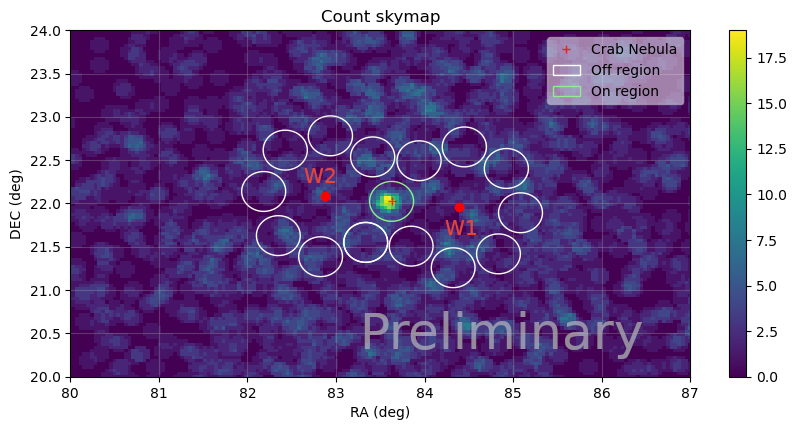}
\end{tabular}
\caption{Smoothed count map of the events recorded during 2h38m of Crab nebula observation. The red points show the two wobble positions W1 and W2. White circles show the off regions with 0.23° radius. Red cross shows the sky position of the crab nebula.}
\label{fig.cmap}
\end{figure}

The observation strategy employed during the SST-1M Crab observation involved the utilization of the wobble pointing technique \cite{fomin_new_1994} with two distinct wobble positions, each offset from the Crab nebula by 0.7 degrees.
For each wobble position, the hadronic background is estimated on seven off regions evenly distributed around the camera center. The wobble positions and the on and off regions are shown in Figure \ref{fig.cmap}.

The total duration of the Crab observation campaign exceeded 10 hours.
%,  weather conditions presented the main limitation. 
To optimize data quality, selection criteria were applied, including restricting runs to zenithal angles below 45°, exclusively retaining stereo observations, and eliminating runs with pointing errors. As a result, the final dataset consists of observations obtained within a zenithal angle range of 35° to 45°. The dataset represents a total duration of 2 hours and 38 minutes, with 1.8 hours acquired for the first wobble position and 0.8 hours acquired for the second wobble position.

\subsection{Analysis steps}
\label{ana}

The analysis of the SST-1M data is conducted utilizing \texttt{sst1mpipe}, a dedicated pipeline built upon \texttt{ctapipe} \cite{ctapipe-icrc-2021}, a framework designed for prototyping and implementing low-level data processing algorithms for CTA.  The details of the analysis pipeline is described in \cite{jurysek_sst1m_2023}.

The first step is to subtract the baseline and apply the calibration described in section \ref{calib}. The signal is then integrated over a 28 ns window starting 12 ns before the maximum of the light pulse for each pixel. Resulting images are cleaned using a tailcut cleaning method which involves applying two thresholds (8 p.e. and 4 p.e.) and a minimum of two neighboring pixels above the second threshold to be kept in the final image. The matching of stereo events, based on their local timestamps, is performed after the cleaning process.
Following the Hillas parametrization of the cleaned images, the reconstruction of gamma-ray energy and direction is performed. This reconstruction rely on random forest algorithms trained on Monte-Carlo simulations, as detailed in \cite{jurysek_sst1m_2023}. Furthermore, the gammaness of each event is evaluated through a random forest classifier trained on Monte-Carlo simulations of diffuse protons and gamma-rays. This gammaness estimation allows the effective discrimination between gamma-ray events and background noise from other cosmic-ray particles.

\subsection{Results}
In this section, we present the preliminary results from the stereo observations of the Crab Nebula using the SST-1M telescopes. A gammaness cut of 0.8 was applied based on Monte Carlo studies to optimize the signal-to-noise ratio. The analysis utilized a theta square cut of 0.02 deg$^2$, corresponding to a 0.14° 68\% containment of the gamma psf function at a zenithal angle of 40°.

The analysis resulted in a significant detection of the Crab nebula, with an excess of 18 events leading to a Li\&Ma significance of 5.21 sigma. The expected excess, determined through Monte Carlo simulations incorporating both point-like gamma-rays rescaled to match the Crab spectrum \cite{ALEKSIC201676} and cosmic rays \cite{DeMitri:2021jog}, is estimated to be 28 ± 5 for a total observation time of 2 hours and 38 minutes.

\begin{figure}[h!]
\centering
\begin{tabular}{c}
\includegraphics[width=.8\textwidth]{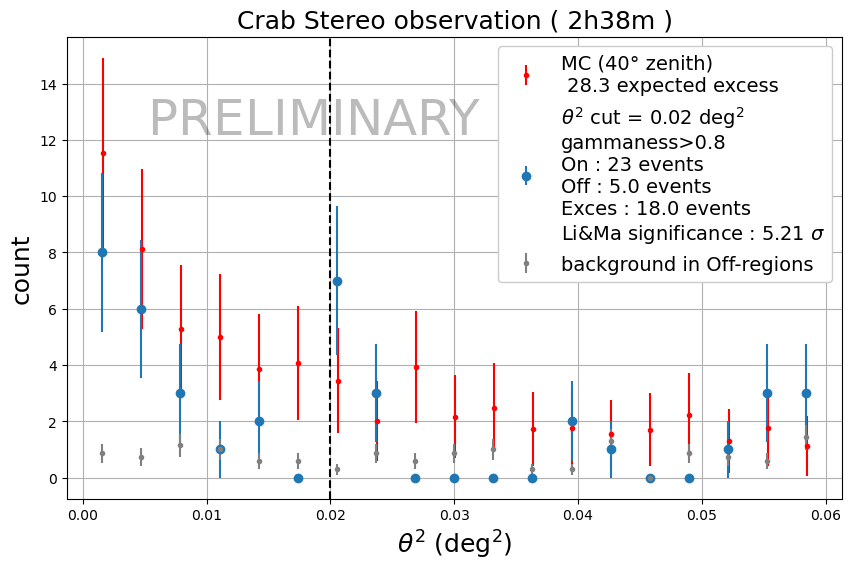}
\end{tabular}
\caption{Theta square distribution of events accumulated during 2.63 hours of the Crab nebula in March 2023 (blue points). Red points show the expected distribution estimated through Monte Carlo simulations including both point-like gamma-rays rescaled to match the Crab spectrum (\cite{ALEKSIC201676}) and cosmic rays (\cite{DeMitri:2021jog}) for the same observation time. Grey points are the Cosmic ray background estimated on the 7+7 off-regions.}
\label{fig:theta2}
\end{figure}

The obtained results show a good agreement between the expected performance of the instrument presented in \cite{jurysek_sst1m_2023} and its actual performance, taking into account a potential discrepancy in energy thresholds between the real instrument and its Monte-Carlo model, as discussed in Section \ref{muon}. This agreement confirms the reliability and accuracy of the instrument's design and validates the effectiveness of the implemented calibration and analysis pipeline.

The $\theta^2$ distribution for both data and Monte-Carlo simulation are shown in Figure \ref{fig:theta2}.

\section{Conclusion}
We presented an overview of SST-1M analysis pipeline from the calibration to preliminary results of the Crab nebula observation.
The calibration of the instrument, including the determination of the gain and crosstalk probability, was achieved through on-site measurements and analysis of multiple photo-electron spectra obtained during dedicated dark runs.
The analysis of cosmic muon images provided valuable information on the optical throughput of the instrument, contributing to the refinement of future Monte Carlo simulations.

The stereo observations of the Crab nebula with the SST-1M telescopes resulted in a significant detection within a data-taking duration of 2 hours and 38 minutes. The observed agreement between the instrument's expected performance and its actual performance validates the instrument's design, calibration procedures, and analysis pipeline.
These preliminary results mark a significant milestone and lay a promising foundation for the development of a stereo analysis pipeline for the SST-1M telescopes and future observations.

%%%%%%%%%%%%%%
\section*{Acknowledgements}\footnotesize
The contribution of the Czech authors is supported by research infrastructure CTA-CZ, LM2023047 MEYS, and Czech Science Foundation, GACR 23-05827S. Funding by the Polish Ministry of Education and Science under project DIR/WK/2017/2022/12-3 is gratefully acknowledged. The contribution of the Départment de Physique Nucléaire et Corpusculare, Faculty de Sciences of the University of Geneva,  1205 Genève is supported by The Foundation Ernest Boninchi,1246 Corsier-CH, and the Swiss National Foundation (166913, 154221, 150779, 143830).

%%%%%%%%%%%%%%

%%%%%%%%%%%%%%
\bibliographystyle{JHEPe}
{\footnotesize
\setlength{\bibsep}{0.2pt}
\bibliography{references}}

\end{document}